# Tuning electronic structures via epitaxial strain in $Sr_2IrO_4$ thin films

J. Nichols, J. Terzic, E. G. Bittle, O. B. Korneta, L. E. De Long, J. W. Brill, G. Cao, and S. S. A. Seo[a)]

*Department of Physics and Astronomy, University of Kentucky, Lexington, KY 40506, USA*

## Abstract

We have synthesized epitaxial $Sr_2IrO_4$ thin-films on various substrates and studied their electronic structures as a function of lattice-strain. Under tensile (compressive) strain, increased (decreased) Ir-O-Ir bond-angle is expected to result in increased (decreased) electronic bandwidth. However, we have observed that the two optical absorption peaks near 0.5 eV and 1.0 eV are shifted to higher (lower) energies under tensile (compressive) strain, indicating that the electronic-correlation energy is also affected by in-plane lattice-strain. The effective tuning of electronic structures under lattice-modification provides an important insight into the physics driven by the coexisting strong spin-orbit coupling and electronic correlation.

[a)] E-mail: a.seo@uky.edu

Recent studies of 5*d* transition-metal oxides have revealed innovative physics with strong potential for electronic device applications. Although weak electronic correlations and paramagnetism were conventionally anticipated to govern the physics of 5*d* transition-metal oxides owing to the extended nature of 5*d* electrons, recent research on $Sr_2IrO_4$ crystals has shown that $J_{eff} = 1/2$ Mott states appear due to strong spin-orbit interactions.[1] The energy scale of the spin-orbit coupling is comparable to the crystal-field energy and the on-site Coulomb interaction, which creates the potential for the emergence of unprecedented electronic states due to the strong competition between these fundamental interactions. Moreover, strong spin-orbit interactions can also derive topologically protected electronic states,[2] the so-called *topological insulator*.

To understand the physics of $Sr_2IrO_4$ and to find a way of tuning its multiple, competing interactions, experimental studies of chemical doping[3-9] and high pressure[10] have been recently conducted. However, chemical doping often has multiple consequences, such as changes in crystal structure and the valence states of the transition metal element, which can make it difficult to achieve a desired state. Hydrostatic pressure techniques are not available for a few advanced spectroscopic methods such as X-ray photoemission spectroscopy. Hence, synthesis of epitaxial thin-films under various lattice-strain conditions is a viable alternative for creating desired modifications of the physical properties of $Sr_2IrO_4$, and there have been only a few thin-film studies.[9,11]

In this letter, we report on the growth and optical properties of $Sr_2IrO_4$ (SIO-214) thin films. The in-plane lattice mismatches between SIO-214 and various oxide substrates can exert both tensile (+) and compressive (–) strains to films, as shown in Fig. 1(a). We find that the electronic structure of SIO-214 films are effectively altered by lattice strain, and we observe shifted optical transitions (absorptions) between the $J_{eff}$ = 1/2 lower Hubbard band (LHB) and the $J_{eff}$ = 1/2 upper Hubbard band (UHB), and between the $J_{eff}$ = 3/2 band and the $J_{eff}$ = 1/2 UHB band. Our observations strongly suggest that not only the electronic bandwidth, but also the magnitude of the effective electronic correlation energy ($U_{eff}$), can be manipulated by lattice strain. Our results demonstrate that epitaxial SIO-214 thin films can be used as a model system to study the physics of coexisting strong electron correlation and strong spin-orbit coupling under lattice modification.

We have used a custom-built, pulsed laser deposition system equipped with *in-situ* reflection high-energy electron diffraction and optical spectroscopic ellipsometry[12] to grow epitaxial SIO-214 thin films 10 unit-cells thick (approx. 25 nm in thickness). Five different single-crystalline substrates are used: $GdScO_3$ (110) (GSO), $SrTiO_3$ (100) (STO), $(LaAlO_3)_{0.3}(Sr_2AlTaO_6)_{0.7}$ (100) (LSAT), $NdGaO_3$ (110) (NGO), and $LaAlO_3$ (110) (LAO). Note that the [110] surface-normal directions of orthorhombic substrates provide in-plane square lattices ($a_{[1\text{-}10]} \approx a_{[001]}$), as listed in Table I. The epitaxial growth conditions are found to be the following: an oxygen partial pressure ($P_{O2}$) of 10 mTorr, a substrate temperature of 700 °C, and a laser (KrF excimer, $\lambda$ = 248 nm) fluence of 1.2 J/cm$^2$. Figure 2 shows θ-2θ X-ray diffraction scans of the samples discussed herein. Well-defined 00*l*-peaks are present due to the films' 00*l*-orientation along the perpendicular to the substrates. The full widths at half maximum in rocking-curve scans of the 00*l* peaks are all less than 0.05°, which confirms the high crystallinity of the films. Note that the thin films' 00$\underline{12}$-peaks are shifted to low angles as the substrate lattice parameters decrease (from GSO to LAO). This behavior is consistent with the schematic diagrams in Fig. 1(b), since elongated (contracted) out-of-plane lattice parameters are expected as compressive (tensile) in-plane strain is exerted on thin films.

Figure 3(a) shows X-ray reciprocal space maps, which reveal important information about both the in-plane and the out-of-plane lattice parameters of the SIO-214 thin films near the 332-reflection (103-reflection) of orthorhombic (pseudo-cubic) substrates. The $11\underline{18}$-peaks from the thin films are clearly observed, and are shifted to low reciprocal wave vectors ($Q_\perp$) as the lattice parameters of the substrate decrease from GSO to LAO, which is also consistent with the $\theta$-$2\theta$ scan data shown in Fig. 2. Note that the in-plane lattice parameters of the films are coherently strained to the substrates when there are small lattice-mismatches (STO and LSAT substrates), or partially-relaxed when the lattice mismatches are relatively large (GSO, NGO, and LAO substrates). Figure 3(b) shows that the in-plane and out-of-plane lattice parameters change systematically as the substrate lattice parameters vary. From this data, we can calculate the in-plane lattice strain ($\varepsilon_{xx}$), the out-of-plane strain ($\varepsilon_{zz}$), and the Poisson's ratio ($v$) of the films, as summarized in Table I.

To investigate the electronic structure of the SIO-214 thin-films, we measured optical transmission spectra ($T(\omega)$) to obtain the optical absorption spectra, $A(\omega) = -\ln[T(\omega)_{\text{film+substrate}} / T(\omega)_{\text{substrate}}]/t$, where $t$ is the thickness of the thin films. Each spectrum has been measured at room temperature using a Fourier-transform infrared spectrometer in the photon energy region of 50 meV – 0.6 eV and a grating-type spectrophotometer in the range of 0.5 – 6 eV. Figure 4 shows the optical absorption spectra of the SIO-thin films grown on GSO, STO, LSAT, and LAO substrates. Spectra of the films grown on NGO substrates cannot be obtained by our method since NGO substrates are opaque in this photon-energy region.

The lattice-strain-dependent optical absorption spectra yield indispensable information about the electronic structure of the SIO-214 thin films. There are two optical absorption peaks at about 0.5 eV ($\alpha$) and 1.0 eV ($\beta$), which have been interpreted as interband optical transitions from the LHB of $J_{eff} = 1/2$ and the $J_{eff} = 3/2$ bands to the unoccupied UHB of the $J_{eff} = 1/2$ state, respectively.[13,14] The interband optical transition intensity, $I_{i \to f}(\omega_0)$, from an initial state $i$ to a final state $f$ at $\omega_0$ is described as $I_{i \to f}(\omega_0) = \int |\langle f|M|i\rangle|^2 \rho_f(\omega)\rho_i(\omega - \omega_0)d\omega$ according to Fermi's golden rule, where $M$ is the matrix element, and $\rho_i$ and $\rho_f$ are the densities of states for $i$ and $f$, respectively. We can expect these optical peaks to broaden (narrow) as the bond angle (Ir-O-Ir) of the in-plane lattice increases (decreases), which enhances (reduces) the electronic hopping integral and the bandwidth (W). When tensile (compressive) strain is applied to SIO-214 thin films, the optical peak width clearly increases (decreases), as shown in Fig. 4. This behavior is consistent with an increased bond angle and an unaffected planar Ir-O bond length, which is very difficult to alter, as schematically shown in Fig. 1(b). In addition, it has been recently shown that for SIO-214 thin films grown on STO, that in-plane tensile strain only affects the bond angle.[11] Surprisingly, our data (Figure 5(a)) shows that not only the optical peak width but also the peak energies ($\omega_\alpha$ and $\omega_\beta$) increase as tensile strain is applied to the SIO-214 thin films. Since both $\omega_\alpha$ and $\omega_\beta$ vary with applied strain, it is unlikely that the spin-orbit coupling can be solely responsible for these changes in the peak energy. This means that $U_{eff}$ should increase (decrease) as tensile- (compressive) strain is applied to these films. As a result, the SIO-214 films exhibit optical gap energies (~0.3 eV) that are approximately independent of lattice strain, as shown in Fig. 4.

This behavior of the optical transitions under lattice-strain is quite unexpected. Temperature-dependent optical measurements[15] on bulk single-crystal $Sr_2IrO_4$ shows that the widths of the optical peaks increase as temperature increases, since the effective Ir-O-Ir bond-angle increases due to thermal expansion, which is qualitatively consistent with our observations in the SIO-214 thin films. However, the temperature-dependent data for bulk crystal shows that the optical peak energies decrease as temperature increases, which is an opposite behavior to our observations in thin films, in which optical peak energies increase under tensile strain. Whereas the thermal effect should increase the overall lattice parameters of a sample, our thin films under biaxial in-plane tensile-strain expand only the in-plane lattice parameters, but reduce the out-of-plane lattice parameters. Hence, the apparent inconsistency between the temperature-dependent experiment and our strain-dependent experiment might not be so surprising. However, in general, theoretical models about 214 structures assume ideal in-plane square lattices and neglect changes along the out-of-plane directions. Hence, our observations tell us that the change along the out-of-plane direction such as apical Ir-O bond-length or interlayer coupling cannot be neglected when considering the actual electronic structure of SIO-214. Figure 5(b) describes qualitatively the change in the electronic structure of the strongly correlated, spin-orbit-coupled bands of SIO-214 thin films under lattice strain. In order to further clarify this change in the electronic structure of the SIO-214 thin films, advanced spectroscopic techniques, such as angle-resolved photoemission spectroscopy and resonant X-ray scattering, should be pursued.

In summary, epitaxial $Sr_2IrO_4$ thin films have been grown on various substrates for both tensile- and compressive-lattice strain. We have observed that optical absorption peaks at around 0.5 eV and 1.0 eV are broadened and shifted to higher energy as tensile strain is applied. This implies that not only is the electronic bandwidth increased by strain, but the electronic correlation is increased as well. These epitaxial $Sr_2IrO_4$ thin films provide a useful two-dimensional model system to study the nature of the coexistence of strong electron correlation and spin-orbit interactions, and their behavior under lattice modifications.

We thank B. I. Min, H.-Y. Kee, G. Jackeli, S. J. Moon, D. Haskel, and B. J. Kim for insightful discussions. This research was supported by the NSF through Grant No. EPS-0814194 (the Center for Advanced Materials), Grant No. DMR-0800367, Grant No. DMR-0856234, by U.S. DoE through Grant No. DE-FG02-97ER45653, and by the Kentucky Science and Engineering Foundation with the Kentucky Science and Technology Corporation through Grant Agreement No. KSEF-148-502-12-303.

**Table I**. In-plane lattice mismatches and corresponding strains of $Sr_2IrO_4$ (SIO-214) films ($t$ ~ 25 nm).

| Substrate | Pseudo-cubic in-plane lattice, $a_{[1\text{-}10]} \approx a_{[001]}$ (Å) | Lattice mismatch (%) [a] | $\varepsilon_{xx}$ (%) [b] | $\varepsilon_{zz}$ (%) [b] | Poisson's ratio [c] |
|---|---|---|---|---|---|
| GSO (110) | 3.96 | +1.53 | +1.64 (±0.44) | –1.69 (±0.09) | 0.34 (±0.08) |
| STO (100) | 3.90 | +0.45 | +0.38 (±0.13) | –0.37 (±0.13) | 0.33 (±0.16) |
| LSAT (100) | 3.87 | –0.45 | +0.26 (±0.24) | +0.00 (±0.10) | 0.00 (±0.18) |
| NGO (110) | 3.81 | –2.04 | –1.82 (±0.08) | +0.39 (±0.12) | 0.09 (±0.02) |
| LAO (110) | 3.79 | –2.58 | –1.18 (±0.37) | +0.67 (±0.10) | 0.22 (±0.04) |

[a] Values are calculated from the pseudo-cubic lattice parameters of bulk SIO-214 ($a_{bulk}$) and substrates ($a_{sub}$) by $(a_{sub} - a_{bulk}) / a_{sub} \times 100$ (%).

[b] Values are obtained using $\varepsilon_{xx} = (a_{film} - a_{bulk}) / a_{bulk} \times 100$ (%) and $\varepsilon_{zz} = (c_{film} - c_{bulk}) / c_{bulk} \times 100$ (%).

[c] Poisson's ratio, $v = \varepsilon_{zz} / (\varepsilon_{zz} - 2\,\varepsilon_{xx})$.

**Figure captions**

FIG. 1. (a) Comparison of pseudo-cubic, in-plane lattice parameter of $Sr_2IrO_4$ with various oxide substrates. (b) Schematic diagrams of relations between lattice strain and Ir-O-Ir bond-angle ($\theta$).

FIG. 2. X-ray θ-2θ diffraction scans of SIO-214 films grown on GSO, STO, LSAT, NGO, and LAO (from top to bottom). 00*l* peaks from the SIO-214 films ($l$ = 4, 8, 12, 16, 24, 28) are clearly observed. The left panel shows zoomed-in scans near the 00<u>12</u>-reflections of the films and the pseudo-cubic 002-reflections of the substrates. The dashed line indicates the 00<u>12</u>-peak position of SIO-214 single crystals. The asterisks (*) indicate peaks from the substrates.

FIG. 3. (a) X-ray reciprocal space maps ($Q_{\parallel} \equiv 2\pi/d_{\parallel}$, $Q_{\perp} \equiv 2\pi/d_{\perp}$) around the 332-reflection (or 103-reflection in pseudo-cubic notation) of the substrates, where $d_{\parallel}$ and $d_{\perp}$ are in-plane and out-of-plane lattices, respectively. While the in-plane lattices of SIO-214 films are coherently strained with respect to the STO and LSAT substrates, partial strain-relaxations are visible from the films grown on the GSO, NGO, and LAO substrates. (b) In-plane ($a$) and out-of-plane ($c$) lattice parameters of the SIO-214 thin films obtained from X-ray scans are plotted as a function of substrate lattice parameters. The dashed and dotted lines indicate the out-of-plane and the in-plane lattice parameters, respectively, of a SIO-214 single crystal.

FIG. 4. Optical absorption spectra of SIO-214 thin films grown on GSO, STO, LSAT, and LAO substrates. The spectra are shifted for clarity. The solid lines are fit curves using two Lorentz oscillators for peaks α and β with a background oscillator. The dashed lines show the estimated optical gap energies (~ 0.3 eV) for the SIO-214 thin films.

FIG. 5. (a) The peak positions ($\omega_{peak}$) and the widths ($\gamma_{peak}$) obtained from Lorentz oscillator fitting as a function of the in-plane lattice strain ($\varepsilon_{xx}$). The values at $\varepsilon_{xx} = 0$ are taken from a SIO-214 single crystal.[13] (b) Schematic band diagrams dependence on lattice strain suggested from our data.

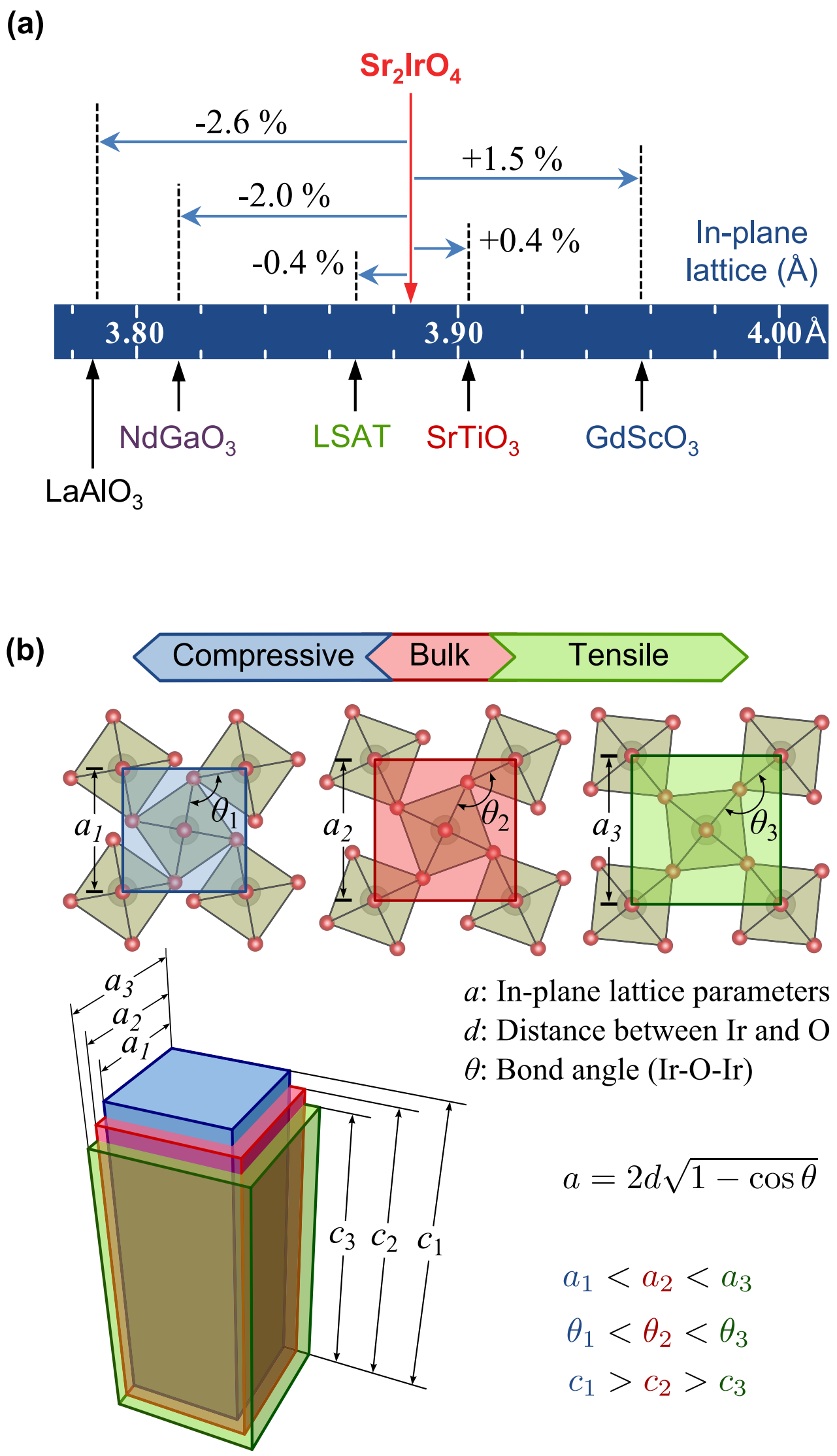

Nichols *et al.*
Figure 1

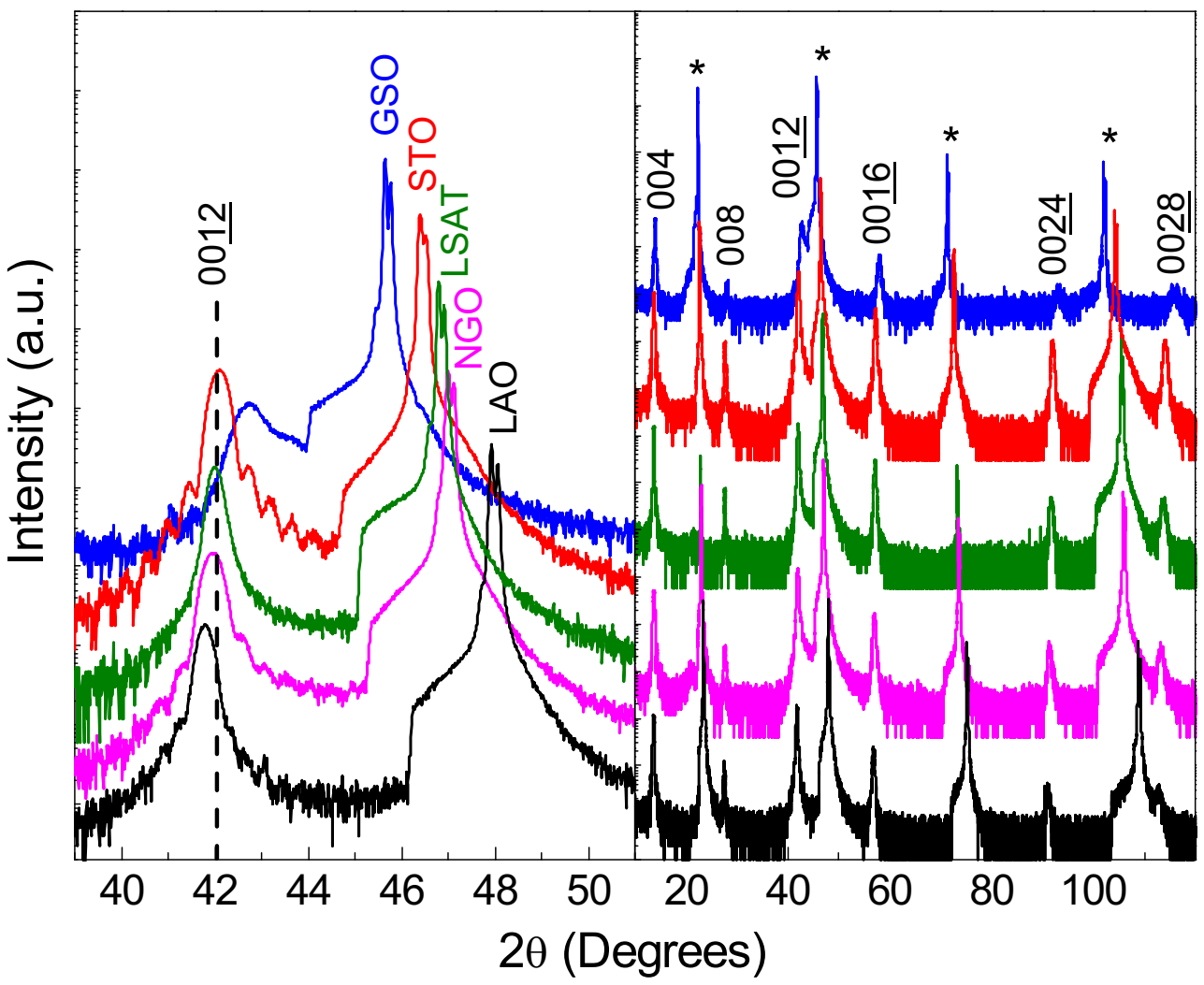

Nichols *et al.*
Figure 2

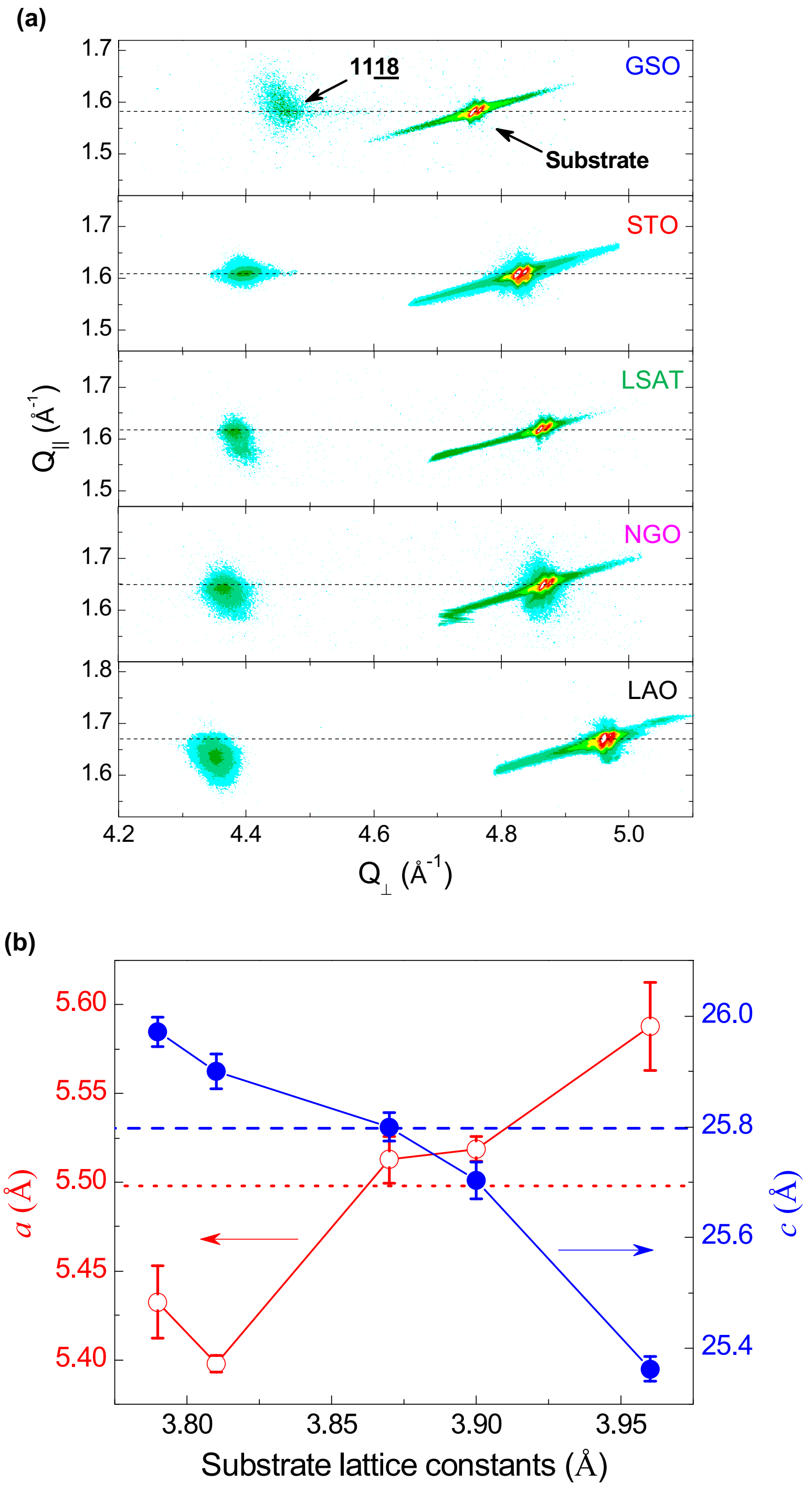

Nichols *et al.*
Figure 3

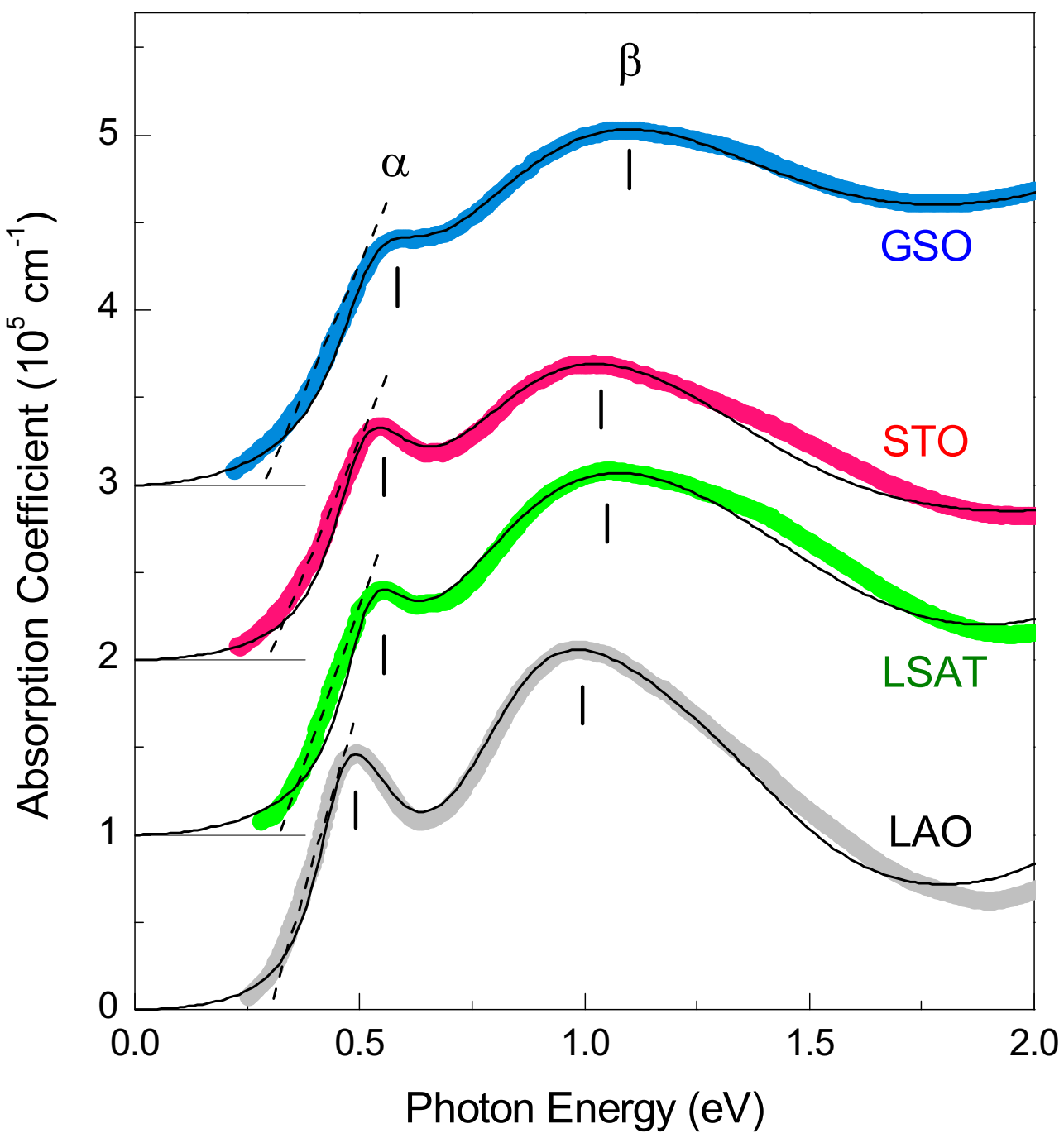

Nichols *et al.*
Figure 4

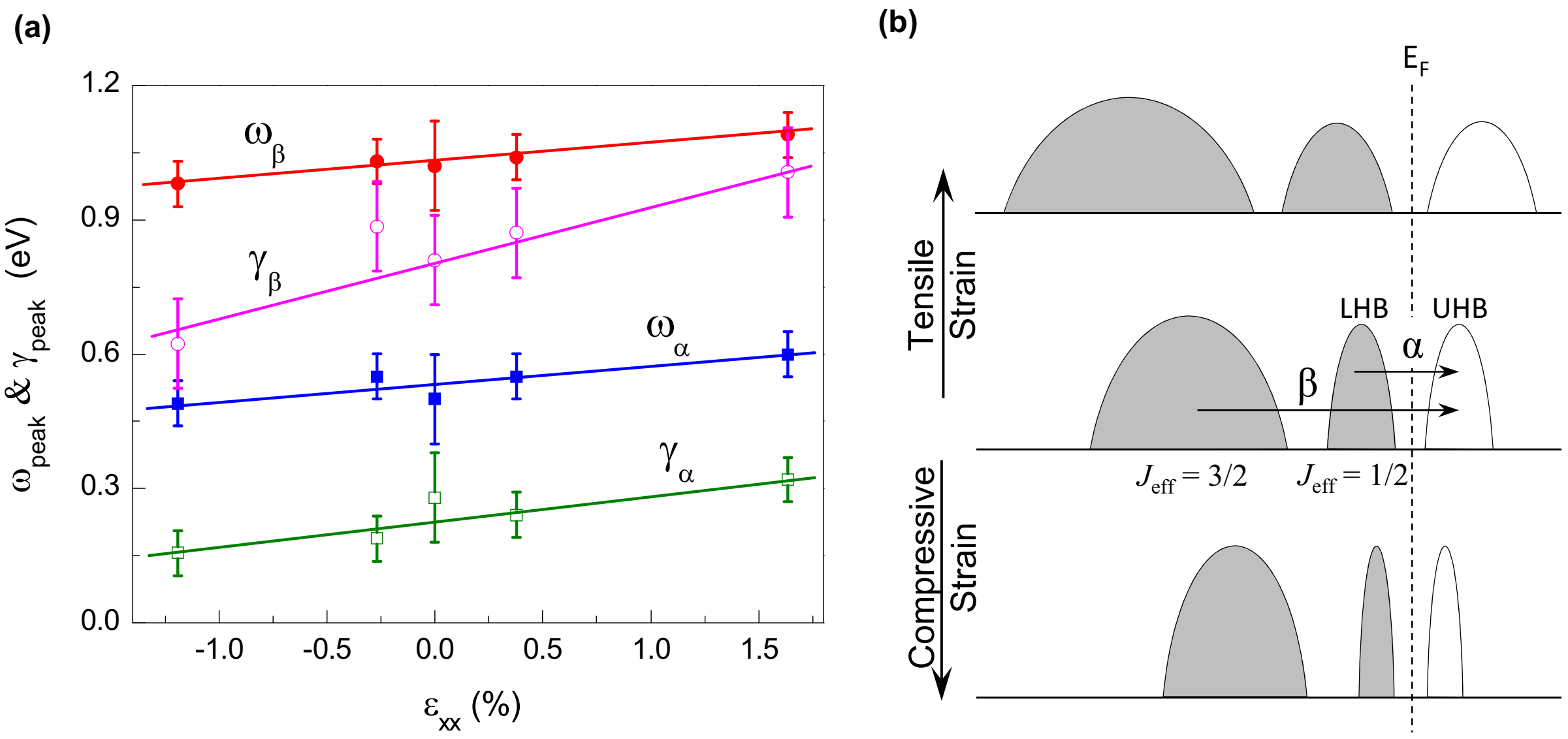

Nichols *et al.*
Figure 5